\documentclass{journal}
\usepackage[super,sort&compress,comma]{natbib}
\usepackage[utf8]{inputenc}
\usepackage[english]{babel}
\usepackage{graphicx}
\usepackage{hyperref}
\usepackage{amssymb}
\usepackage{mathtools}
\usepackage{xr}
\usepackage{booktabs}
\usepackage{xcolor}

\DeclareFontFamily{U} {MnSymbolC}{}
\DeclareFontShape{U}{MnSymbolC}{m}{n}{
  <-6> MnSymbolC5
  <6-7> MnSymbolC6
  <7-8> MnSymbolC7
  <8-9> MnSymbolC8
  <9-10> MnSymbolC9
  <10-12> MnSymbolC10
  <12-> MnSymbolC12}{}
\DeclareFontShape{U}{MnSymbolC}{b}{n}{
  <-6> MnSymbolC-Bold5
  <6-7> MnSymbolC-Bold6
  <7-8> MnSymbolC-Bold7
  <8-9> MnSymbolC-Bold8
  <9-10> MnSymbolC-Bold9
  <10-12> MnSymbolC-Bold10
  <12-> MnSymbolC-Bold12}{}
\DeclareSymbolFont{MnSyC} {U} {MnSymbolC}{m}{n}
\DeclareMathSymbol{\filledstar}{\mathbin}{MnSyC}{129}
\DeclareMathSymbol{\medbullet}{\mathbin}{MnSyC}{89}

\hypersetup{
    colorlinks=true,
    linkcolor=blue,
    citecolor=blue,
    urlcolor=blue,
}

\DeclareMathOperator\erfc{erfc}

\begin{document}
\articletype{Paper}

\title{XUV fluorescence as a probe of interatomic Coulombic decay of resonantly excited He nanodroplets}
\author{
  Keshav Sishodia,\textit{$^{1, 2}$}
  Ltaief Ben Ltaief,\textit{$^{3}$}
  Niklas Scheel,\textit{$^{3}$}
  Istv{\'a}n B. F{\"o}ldes,\textit{$^{4}$}
  Andreas Hult Roos,\textit{$^{2}$}
  Martin Albrecht,\textit{$^{2}$}
  Maty\'{a}\v{s} Stan\v{e}k,\textit{$^{2, 5}$}
  Lucie Jurkovi\v{c}ov\'{a},\textit{$^{2, 5}$}
  Ondrej Hort,\textit{$^{2}$}
  Jaroslav Nejdl,\textit{$^{2, 5}$}
  Ernesto Garc{\'i}a-Alfonso,\textit{$^{6}$}
  Nadine Halberstadt,\textit{$^{6}$}
  Jakob Andreasson,\textit{$^{2}$}
  Eva Klime\v{s}ov\'{a},\textit{$^{2}$}
  Maria Krikunova,\textit{$^{2, 7}$}
  Sivarama Krishnan,\textit{$^{1}$}
  Andreas Heidenreich,\textit{$^{8, 9}$}
  and Marcel Mudrich$^{\ast}$\textit{$^{10}$}
}

\affil{$^{1}$~Department of Physics and Quantum Center of Excellence for Diamond and Emergent Materials, Indian Institute of Technology Madras, 600036 Chennai, India}

\affil{$^{2}$~ELI Beamlines facility, The Extreme Light Infrastructure ERIC, Za Radnic\'{\i} 835, 253 41 Doln\'{\i} B\v{r}e\v{z}any, Czech Republic}

\affil{$^{3}$~Department of Physics and Astronomy, Aarhus University, Ny Munkegade 120, 8000 Aarhus C, Denmark}

\affil{$^{4}$~HUN-REN Wigner Research Centre for Physics, H-1121 Budapest, Hungary}

\affil{$^{5}$~Czech Technical University in Prague, FNSPE, 115 19 Prague, Czech Republic}

\affil{$^{6}$~Laboratoire Collisions, Agr{\'e}gats, R{\'e}activit{\'e} (LCAR), Université de Toulouse, CNRS, 31062 Toulouse, France}

\affil{$^{7}$~Waldau Technical University of Applied Sciences, Hochschulring 1, 15745 Wildau, Germany}

\affil{$^{8}$~Kimika Fakultatea, Euskal Herriko Unibertsitatea (UPV/EHU) and Donostia International Physics Center (DIPC), P.K. 1072, 20080 Donostia, Spain}

\affil{$^{9}$~IKERBASQUE, Basque Foundation for Science, 48011 Bilbao, Spain}

\affil{$^{10}$~Institute of Physics, University of Kassel, 34132 Kassel, Germany}

\email{mudrich@phys.au.dk}

\begin{abstract}
Superfluid He nanodroplets resonantly excited by extreme ultraviolet (XUV) pulses exhibit complex relaxation dynamics, including the formation of metastable excited He$^*$ atoms trapped in bubbles, the desorption of excited atoms from the droplet surface, and autoionization via interatomic Coulombic decay (ICD). Irradiation with intense infrared pulses can trigger avalanche ionization, leading to the formation and subsequent expansion of a He nanoplasma. Here, we introduce a novel approach to probe the ICD dynamics over timescales spanning femtoseconds to nanoseconds. Our method exploits the efficient ignition of a nanoplasma through tunnel ionization of excited helium atoms attached to the droplets and the detection of XUV fluorescence emitted from the resulting nanoplasma. Using quantum mechanical and classical calculations, we interpret the nanosecond fluorescence decay as a signature of ICD mediated by He$^*$ freely roaming on the nanodroplet surface.
\end{abstract}

\section{Introduction}
\label{sec:intro}

Helium (He) nanodroplets are intriguing quantum fluid clusters that exhibit unique characteristics distinct from other atomic and molecular clusters.
The He atoms within these droplets are weakly bonded through minimal attractive London dispersion forces.
Due to their low mass, they possess significant zero-point energy, leading to collective quantum effects.
Remarkably, the He nanodroplets cool evaporatively to an ultracold temperature of 0.37~K, where microscopic superfluidity arises.
\cite{grebenevSuperfluiditySmallHelium41998,tangQuantumSolvationCarbonyl2002,brauerCriticalLandauVelocity2013}
Despite their inert nature, excitation or ionization can transform He nanodroplets into a highly reactive medium, facilitating a variety of interatomic and intermolecular processes.
\cite{mudrichPhotoionisatonPureDoped2014,laforgeInteratomicIntermolecularDecay2024}

Resonant excitation of He nanodroplets initiates an intricate coupled electronic and nuclear relaxation dynamics.
\cite{mudrichUltrafastRelaxationPhotoexcited2020,asmussenUnravellingFullRelaxation2021,laforgeRelaxationDynamicsExcited2022}
After localization of the excitation on a metastable He$^*$ atom, within $\lesssim1$~ps, a void cavity (`bubble') with radius $R=0.68$~nm forms around the He$^*$ atom.
\cite{vonhaeftenObservationAtomiclikeElectronic2001,closserInitioCalculationsElectronically2010,kornilovFemtosecondPhotoelectronImaging2011,asmussenUnravellingFullRelaxation2021}
The repulsive interaction between the He$^*$ and the surrounding He atoms leads to a massive broadening of the spectral lines of He nanodroplets in the absorption spectra.
\cite{joppienElectronicExcitationsLiquid1993}
However, these absorption bands largely retain their atomic character.
The localized He$^*$ subsequently emerges to the droplet surface from which it is either ejected into the vacuum or remains loosely bound and roams about the droplet surface.
The first time-resolved experiments on pure He nanodroplets seemed to indicate that He$^*$ excited atoms are readily ejected from the droplets.
\cite{kornilovFemtosecondPhotoelectronImaging2011}
However, subsequent measurements showed that the ejection of He$^*$ atoms is actually a minor relaxation channel.
\cite{ziemkiewiczFemtosecondTimeresolvedXUV2014,ziemkiewiczUltrafastElectronicDynamics2015}
This finding is supported by another observation: Foreign atoms or molecules (`dopants') attached to the He droplets are efficiently ionized by interaction with He$^*$, a process termed resonant interatomic Coulombic decay (rICD) or Penning ionization.
\cite{buchtaChargeTransferPenning2013,asmussenDopantIonizationEfficiency2023,laforgeInteratomicIntermolecularDecay2024}

Upon optical excitation, He$^*$ is formed in a singlet state which predominantly relaxes to the 1s2s\,$^1$S state irrespective of the initial excited state.
\cite{mudrichUltrafastRelaxationPhotoexcited2020,benltaiefChargeExchangeDominates2019,asmussenUnravellingFullRelaxation2021,laforgeRelaxationDynamicsExcited2022}
Direct fluorescence decay from the 1s2p\,$^1$P state back to the ground state is only a weak channel.~\cite{karnbachElectronicExcitationDecay1995}
Triplet states, in particular the 1s2s\,$^3$S metastable state, are populated by electron-ion recombination in the regime of droplet autoionization and electron impact excitation at excitation energies $h\nu>23$~eV and $h\nu>45$~eV, respectively.
\cite{vonhaeftenDiscreteVisibleLuminescence1997,buchtaExtremeUltravioletIonization2013,benltaiefSpectroscopicallyResolvedResonant2024,benltaiefEfficientIndirectInteratomic2023,laforgeInteratomicIntermolecularDecay2024}

The fate of those He$^*$ that relax to the 1s2s\,$^1$S state and remain bound to the droplet surface is not exactly known.
In bulk liquid He, He$_2^*$ excimers form in the $A\,^1\Sigma_u^+$ state by association of the He$^*$($^1$S) with a neighboring ground-state He atom.
This reaction is hindered by a repulsive barrier in the He$^*$-He pair potential.
\cite{buchenauExcitationIonization4He1991,fiedlerInteractionHeliumRydberg2014}
However, it may still occur due to tunneling and many-body interactions,
\cite{nijjarConversionHe23SHe2a3Su2018} leading to the decay of He$_2^*$ within 1.6~$\mu$s by fluorescence emission.
\cite{mckinseyTimeDependenceLiquidhelium2003}
He$_2^*$ excimers can also be formed directly in their lowest vibrational state without crossing the barrier. In that case, they decay by fluorescence emission within $\sim10$~ns.
\cite{mckinseyTimeDependenceLiquidhelium2003}
In He nanodroplets, only a very small fraction of He$^*$ atoms form He$_2^*$ excimers by direct association, as found in earlier time-resolved experiments.
\cite{ziemkiewiczUltrafastElectronicDynamics2015,closserSimulationsDissociationSmall2014}
Likewise, neither photoionization electron spectra~\cite{mudrichUltrafastRelaxationPhotoexcited2020,laforgeRelaxationDynamicsExcited2022} nor rICD electron spectra
\cite{buchtaChargeTransferPenning2013,benltaiefChargeExchangeDominates2019,benltaiefSpectroscopicallyResolvedResonant2024} have shown significant contributions from He$_2^*$ excimers.

The situation changes when multiple He$^*$ are excited in one He nanodroplet at high photon fluence or in large droplets.
In the former case, when a high density of He$^*$ are excited, forming pairs of He$^*$ with short interatomic separations $\lesssim1~$nm, these pairs can decay by homogeneous rICD
\cite{laforgeUltrafastResonantInteratomic2021} according to the reaction
\begin{equation}
\label{eq:He-ICD}
\mathrm{He}^* + \mathrm{He}^* \rightarrow \mathrm{He} + \mathrm{He}^+ + e_\mathrm{ICD}.
\end{equation}
Under these conditions, rICD is an ultrafast process that occurs within $\lesssim1$~ps as it is driven by the He fluid dynamics of the two bubbles forming around the He$^*$ and merging into one.
In the case of low photon fluence and large He droplets, the He$^*$ fully relax and emerge to the droplet surface prior to rICD,
as evidenced by sharp lines in the electron spectra and fully isotropic electron angular distributions.
\cite{benltaiefSpectroscopicallyResolvedResonant2024}
On the one hand, this indicates that this type of rICD of He$^*$ at the droplet surface is a slow process, proceeding on a time scale $\gg 1$~ps after resonant excitation of the He droplet.
On the other hand, the absence of any signature of He$_2^*$ in the rICD electron spectra shows that this rICD proceeds before the formation of He$_2^*$ sets in, \textit{i.\,e.} within $<1.6~\mu$s.
ICD involving He$_2^*$ excimers is observed only in the regimes of autoionization and electron-impact excitation at $h\nu>23$~eV and $h\nu>45$~eV, respectively.
\cite{benltaiefSpectroscopicallyResolvedResonant2024,benltaiefEfficientIndirectInteratomic2023,laforgeInteratomicIntermolecularDecay2024}

In this work, we directly track the rICD process in time under the conditions of low fluence and large He droplets.
He droplets were resonantly excited by a short extreme ultraviolet (XUV) pulse.
To monitor the dynamics of excited He atoms (He$^*$) attached to the droplets,
we introduce a novel probing scheme:
a near-infrared (NIR) laser pulse, delayed with respect to the XUV excitation, is used to avalanche-ionize the droplets and form a nanoplasma.
The resulting fluorescence emission from excited atomic and ionic states is then detected.
These states are populated through avalanche ionization triggered by the NIR pulse, followed by electron-ion recombination.
NIR strong-field ionization of large molecules and clusters is highly sensitive to the presence of free or quasi-free electrons that act as seeds to initiate the ionization avalanche.
Therefore, the fluorescence emission yield is indicative of the population of He$^*$ attached to the He nanodroplets;
He$^*$ are easily tunnel ionized by the intense NIR laser pulse due to their much lower ionization energy ($E_i^\mathrm{He^*}=4$~eV) as compared to ground-state He atoms ($E_i^\mathrm{He}=24.6$~eV). In addition to XUV fluorescence measurements, we have also detected photoelectron spectra from He nanodroplets under the same XUV irradiation conditions to evidence the presence of He$^*$ in the excited nanodroplets and their decay by rICD.

Recently, Sumfleth~\textit{et al.} have measured fluorescence emission spectra from laser-induced He nanoplasmas,
but no time-resolved dynamics were reported.\cite{sumflethXUVFluorescenceProbe2023}
In those experiments, mainly a series of emission lines from excited states of the He$^+$ ion were observed.
Our experiment combines that method with XUV pump and NIR probe time-resolved measurements of the He nanodroplet dynamics.
This approach is similar to an experiment carried out by Medina~\textit{et al.} with a similar setup;
there, the pump-probe dynamics of photoionized He nanodroplets was studied by monitoring the rate of NIR-induced avalanche ionization.
\cite{medinaLonglastingXUVActivation2023}
In that experiment, a long-lasting persistent enhancement of the avalanche ionization probability was found;
in contrast, here we observe a drop of the nanoplasma fluorescence signal on a time scale of about 2-9~ns, with a systematic dependency on the He nanodroplet size. These results are rationalized by quantum mechanical simulations of the structure of He$^*$ binding to the He droplet surface, classical simulations of the motion of He$^*$ atoms at the droplet surface, and classical model calculations of the nanoplasma ignition process. We interpret the observed drop of nanoplasma fluorescence in terms of the decay of pairs of He$^*$ atoms by rICD, mediated by a slow roaming motion of the He$^*$ about the He droplet surface as a consequence of the low temperature and the superfluid nature of He nanodroplets.

\section{Experimental setup}
\label{sec:setup}

\begin{figure}[h]
  \centering
  \includegraphics[width=9cm]{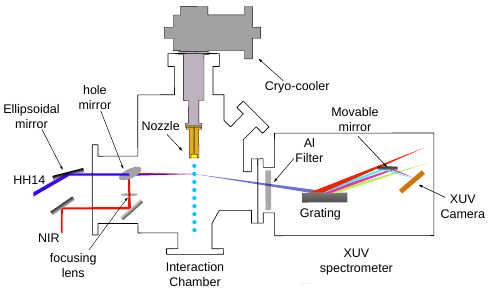}
  \caption{
    \label{fig:setup}
    Schematic representation of the experimental setup.
    The He droplet source is mounted directly into the interaction chamber to take advantage of a high density of He droplets.
    The NIR and XUV beams are focused and superimposed onto the droplet beam in the interaction region.
    The XUV spectrometer is mounted downstream of the interaction region in the direction of propagation of the XUV beam.
  }
\end{figure}

We performed the experiment at the MAC end-station of the high harmonic generation (HHG) Beamline at the Extreme Light Infrastructure (ELI) Beamlines located in Doln\'{i} B\v{r}e\v{z}any, Czech Republic.~\cite{klimesovaMultipurposeEndstationAtomic2021}
Figure~\ref{fig:setup} shows a schematic representation of the experimental setup.
A beam of He nanodroplets was generated by continuous supersonic expansion of He gas at a stagnation pressure of 50 bar out of a cryogenic nozzle with a diameter of 5~$\mu$m cooled to a temperature ranging between 11~K and 14.5~K.
This corresponds to an average size of He droplets between
$\left\langle N \right\rangle = 1\times10^{4}$ and $5\times10^{5}$ He atoms per droplet ($R=5$-17~nm, respectively).
The droplet sizes are determined by comparison with previous measurements .~\cite {toenniesSuperfluidHeliumDroplets2004}
The XUV and NIR pulses were focused in the center of the interaction region at a small angle with respect to each other.

The synchronized XUV and NIR pulses were generated by a commercial femtosecond laser system operating at a repetition rate of 1 kHz.
It provides NIR pulses with energies up to 12 mJ and a pulse duration of $\leq35$~fs centered at a wavelength of 800~nm.
The laser beam was split into an intensity ratio of 8:2, where 80~\% of the beam was guided to the HHG Beamline that generates XUV pulses
\cite{hortHighfluxSourceCoherent2019} and 20~\% of the beam was guided toward the MAC endstation.

In the HHG Beamline, the second harmonic of the NIR beam was generated by passing the beam through a beta-barium borate (BBO) crystal.
The resulting UV beam was focused near the orifice of a pulsed valve from which Kr gas was expanded into the vacuum at a repetition rate of 500~Hz.
The pressure of the Kr gas and the focusing conditions of the second harmonic
beam were optimized to generate the highest intensity of the 14$^{\text{th}}$ harmonic (HH14)
of the NIR at a photon energy of 21.8~eV.~\cite{jurkovicovaBrightContinuouslyTunable2024}
This single harmonic HH14 was selected by a grating monochromator.~\cite{hortHighfluxSourceCoherent2019}
In the interaction region in MAC chamber, the XUV photon flux $\Phi$ was measured using a calibrated XUV-sensitive photodiode,
yielding $\approx$~$5\times10^{6}$~photons/pulse.
The XUV pulse had a pulse duration of $\approx$ 50~fs, an energy bandwidth (FWHM) of 76~meV
and was focused to a spot size (FWHM) of $64\times 47~\mu$m$^{2}$ using an ellipsoidal mirror.

The weaker NIR beam (20~\%) passed through a delay line with a mechanical translation stage that can move up to 1~m with a minimum step size of 20~nm,
allowing us to control the delay between the XUV and the NIR pulses between -0.6~ns and 6~ns.
Due to dispersion in the optical elements, air path of the NIR beam, and partial compression by the chirped mirrors,
the NIR pulse width was about 100~fs in the interaction region.
The pulse energy of the NIR pulse was attenuated to 180-400~$\mu$J.
The NIR pulse was focused to a spot size of $18\times 43~\mu$m$^{2}$ using a focusing lens with a focal length of 150~mm,
yielding a peak intensity of the NIR pulse of $10^{14}$-$10^{15}$~Wcm$^{-2}$.
A mirror with a central bore hole allowed us to recombine the two beams in a nearly collinear geometry in
the interaction region at a distance of about 1~mm from the orifice of the cryogenic nozzle that generated the He nanodroplets.
The positions of the two beam foci were determined by imaging them on a YAG:Ce fluorescence screen
that was inserted into the focal plane after retracting the cryogenic nozzle from the center of the interaction chamber.

An XUV spectrometer was mounted in the direction of propagation of the XUV beam behind the interaction region to detect the fluorescence signal excited by the XUV and NIR beams in He droplets.
The XUV spectrometer consisted of a concave grating with 1200 grooves/mm that dispersed the fluorescence light and focused individual spectral lines.
A selected spectral region was imaged on an XUV camera by a movable mirror.
The XUV spectrometer provides a resolution of 0.05~nm.~\cite{spectroscopyFlatFIeldXUV2020}

To support our investigation using fluorescence spectroscopy, we also measured photoelectron spectra from He droplets under the same conditions in another experimental run.
In that experiment, a magnetic bottle electron spectrometer (MBES) was mounted at the MAC endstation
\cite{elandCompleteTwoElectronSpectra2003, klimesovaUpdateMACEndStation2024} and the droplet beam was skimmed twice to produce a low-density stream of droplets for photoelectron spectroscopy.
The MBES contains a 2~m long flight tube and provides a high electron collection efficiency and an energy resolution of $\Delta E/E \approx 5\%$.~\cite{roosDissociationsWaterIons2018}.
In practice, energy resolution for photoelectron spectroscopy is determined by the energy bandwidth of the XUV pulse.

\section{Results and discussion}
\label{sec:resAndDiscuss}

\subsection{Nanoplasma XUV fluorescence spectra and photoelectron spectra}

\begin{figure}[h]
  \centering
  \includegraphics[width=8.3cm]{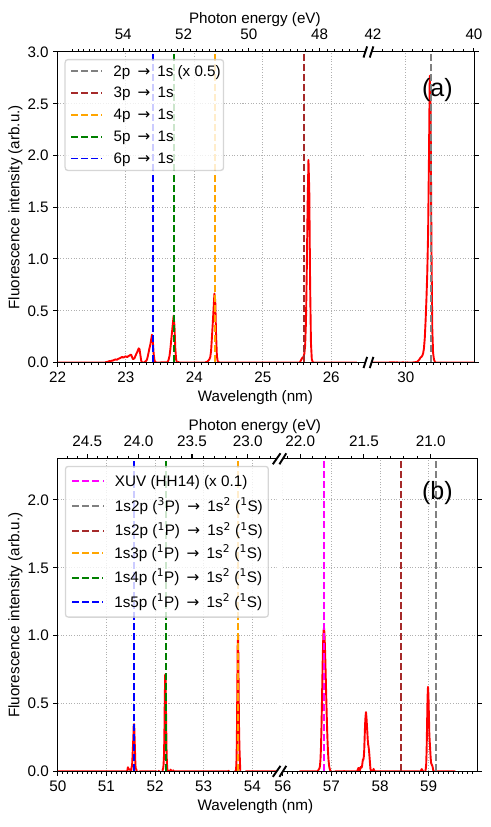}
  \caption{
    \label{fig:fluo}
    Fluorescence spectra emitted from helium nanoplasma containing $3\times 10^{4}$ atoms.
    Droplets were excited by the XUV pulse and
    ignited to the nanoplasma state through avalanche ionization by the temporally delayed NIR pulse with pulse energy of 230$\mu$J.
(a) Spectral lines attributed to excited He ions, He$^{+*}$.
(b) Spectral lines attributed to excited neutral He atoms, He$^{*}$.
The dotted vertical lines show the positions of atomic emission lines from the literature.
The dotted magenta line shows the spectrum of the 14$^{\text{th}}$ harmonic generated in the HHG Beamline.
  }
\end{figure}

Typical spectra of fluorescence emitted by He droplets after ignition of a nanoplasma by the NIR pulse are shown in Fig.~\ref{fig:fluo}.
The strongest fluorescence is observed in the form of emission lines of the excited He ion, He$^{+*}$, see Fig.~\ref{fig:fluo}~(a).
These lines are identified as the transitions 2p$\to$1s at 30.37~nm, 3p$\to$1s at 25.6~nm,
4p$\to$1s at 24.3~nm, 5p$\to$1s at 23.7~nm and 6p$\to$1s at 23.4~nm.~\cite{wieseAccurateAtomicTransition2009}
This observation matches the one by Sumfleth~\textit{et.~al.},~\cite{sumflethXUVFluorescenceProbe2023}
who reported fluorescence spectra in the spectral regions corresponding to the 6p$\to$1s transition and from higher excited states up to the ionization threshold of He$^{+}$ ions for large He droplets containing $1.2\times 10^6$ atoms.
Highly excited He$^{+*}$ cations are populated by recombination of electrons with He$^{2+}$ dications, which subsequently decay by fluorescence.
Additionally, recombination of electrons with He$^+$ monocations in the nanoplasma leads to the formation of excited neutral He$^{*}$ atoms.
The fluorescence emitted by He$^{*}$, shown in Fig.~\ref{fig:fluo}~(b), results from atomic transitions 1s3p $\to$ 1s$^{2}$ at 53.7~nm,
1s4p $\to$ 1s$^{2}$ at 52.2~nm and 1s5p $\to$1s$^{2}$ at 51.5~nm.~\cite{wieseAccurateAtomicTransition2009}
The total fluorescence yield from He$^{+*}$'s was about six times higher than that from He$^{*}$'s,
indicating that a large fraction of He atoms in the nanoplasma are transiently doubly ionized (`inner ionization').
\cite{lastQuasiresonanceIonizationLarge1999}
Subsequent recombination into He$^{+*}$ monocationic states is highly efficient, whereas recombination into neutral He$^{*}$ states remains infrequent, in line with previous observations of the emission of high yields of electrons and He$^+$ ions from the nanoplasma.~\cite{heidenreichChargingDynamicsDopants2017,medinaSingleshotElectronImaging2021}
The fluorescence spectrum of He$^{*}$'s also contains the spectrum of the XUV pump pulse at 56.8~nm.

\begin{figure}[h]
  \centering
  \includegraphics[width=8.3cm]{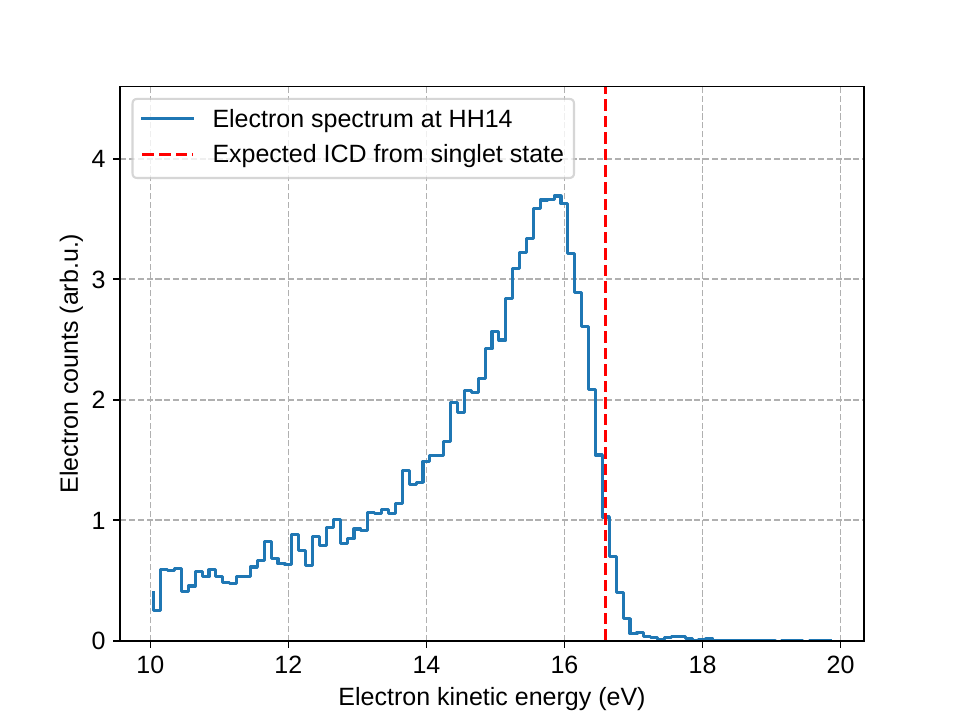}
  \caption{
    \label{fig:ICD}
    Electron spectrum recorded for He nanodroplets containing $5\times 10^5$~He atoms irradiated by a XUV pulse with a photon energy of 21.8~eV. The peak around 15.55~eV manifests the emission of electrons by rICD of multiple He excitations per He droplet. The dashed line indicates the expected electron energy for rICD of pairs of unperturbed He$^*$ atoms in the 1s2s\,$^1$S state.
  }
\end{figure}

The corresponding photon energy of the XUV pump pulse, 21.8~eV, nearly matches the maximum of the main absorption band of He droplets, correlated to the 1s2p\,$^1$P state of He. At this photon energy, the absorption cross section $\sigma$ is estimated at 15~Mb per He atom.
\cite{ovcharenkoNovelCollectiveAutoionization2014,buchtaExtremeUltravioletIonization2013}
The probability of exciting a He atom in a droplet is $p_\mathrm{He} = \Phi \sigma / w^2 \approx 2\times10^{-6}$,
where $w$ is the focal size of the XUV pump pulse.
Consequently, for a droplet with an average radius $R=9$~nm containing $7\times10^4$ He atoms, we anticipate an average of 0.17 He$^*$ excitations per droplet. Therefore, any dynamics due to interacting He$^*$ in one droplet can be safely ascribed to pairs of He$^*$, while the contribution of three or more He$^*$ in one droplet is negligible.

In previous experiments carried out at an XUV free-electron laser (FEL), a higher photon flux led to the excitation of pairs of He$^*$ with short interatomic distances $\lesssim1$~nm;
these pairs decayed by an ultrafast ($\lesssim 1$~ps) rICD process driven mainly by the quantum fluid dynamics of the He surrounding the He$^*$.
\cite{laforgeUltrafastResonantInteratomic2021}
In the current HHG experiment, a similar signature of rICD is observed despite the much lower photon flux.

To unambiguously identify the presence of rICD under our current experimental conditions,
we also measured photoelectron spectra from the droplets irradiated by the XUV pulse.
Figure~\ref{fig:ICD} shows the electron kinetic energy spectrum recorded for He droplets containing an average number of $5\times 10^5$ He atoms irradiated by only the XUV pump pulse.
The peak at 15.55~eV arises from the emission of electrons created by rICD according to Eq.~\ref{eq:He-ICD}, confirming rICD being active under our experimental conditions.
The peak is slightly shifted to lower energy with respect to the expected energy $E_e = 2\times E^* - E_i^\mathrm{He}= 16.6$~eV, taking the atomic term value $E^*=20.6$~eV of the 1s2s\,$^1$S state into account.
This shift and the asymmetric line shape are probably due to elastic scattering of rICD electrons at the He atoms in the droplets.
\cite{asmussenElectronEnergyLoss2023}

In the present case of large droplets and low photon flux, the average distance between the two interacting He$^*$ atoms is much larger as the density of He$^*$ is lower. Thus, ultrafast rICD mediated by the He quantum fluid dynamics is not expected to occur with high probability.
Note that in most FEL experiments, in particular for large He nanodroplets,~\cite{laforgeUltrafastResonantInteratomic2021}
the rICD electron yield at positive delays $\geq 3$~ps did not fully rise up to the level measured at negative delays. This indicated that another slower rICD channel is active in addition to the reported fast rICD process.

\subsection{Time-resolved XUV fluorescence emission}

\begin{figure}[h]
  \centering
  \includegraphics[width=8.3cm]{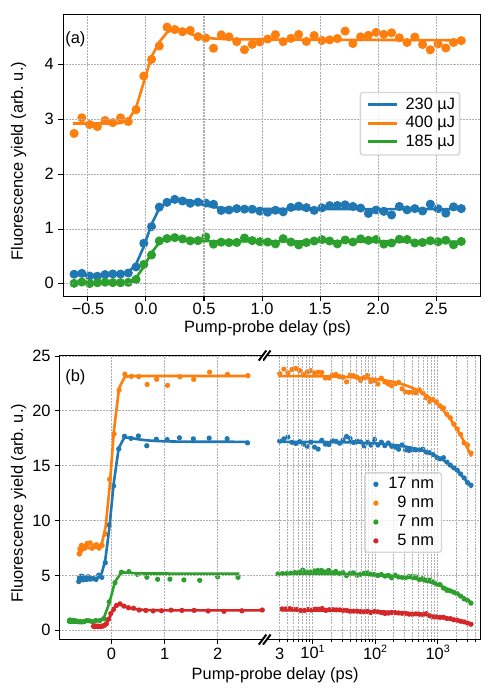}
  \caption{
    \label{fig:pump_probe}
    (a) Fluorescence yield as a function of pump-probe delay
    for different NIR pulse energies for the He droplet with a radius of 7~nm, which contains 30000 He atoms.
    For higher pulse energy, a nanoplasma is produced before the excitation of the droplet by the XUV pulse.
    (b) Fluorescence yield as a function of pump-probe delay
    for different droplet sizes.
    Each curve shows two different decay constants: A fast decay within $< 1$~ps and a slow decay on the ns scale.
    Each curve was fitted by a convolution of two exponential decay functions and the cross-correlation of pump and probe pulses given by Eq.~(\ref{eq:fit_function}).
  }
\end{figure}

Our fluorescence measurements indeed point to the existence of a much slower rICD process that depletes the He$^*$ population on the nanosecond timescale.
Figure~\ref{fig:pump_probe} shows the total fluorescence yield of He$^{+*}$ as a function of the delay between XUV pump and NIR probe pulses.
The colored lines in Fig.~\ref{fig:pump_probe} (a) show total fluorescence yields of He$^{+*}$ for different NIR pulse energies for a He droplet of mean radius $R=7$~nm containing 30,000 atoms.
At negative delay values, when the NIR probe pulse arrives before the XUV pump pulse,
the NIR pulse is inefficient in causing tunnel ionization of He and igniting a nanoplasma when the NIR pulse energy is low ($\leq 230~\mu$J).
Thus, we measure only low yields of nanoplasma fluorescence.
Only at a higher NIR pulse energy of 400~$\mu$J, corresponding to a peak intensity of $\lesssim 10^{15}$~Wcm$^{-2}$ does tunnel ionization occur in the largest He droplets present within the droplet size distribution.
The ignition of the nanoplasma in large He droplets is facilitated by the higher probability for tunnel ionization of one or a few He atoms, as well as the tendency of large He droplets to pick up impurity molecules from the residual gas,
mostly water, which serves as seeds for the NIR-driven avalanche ionization.~\cite{krishnanDopantInducedIgnitionHelium2011}
To systematically investigate the influence of the droplet size distribution on the ignition of nanoplasmas,
we varied the droplet size by changing the nozzle temperature and recorded the delay-dependence of the integrated
He$^{+*}$ fluorescence yield.
A NIR probe pulse with a pulse energy of 230~$\mu$J was set to reach a high pump-probe signal contrast while maintaining a low level of nanoplasma ignition induced by the NIR pulse alone.
Figure \ref{fig:pump_probe}~(b) shows the fluorescence yield as a function of delay for different average droplet sizes.
Note the logarithmic scale at the abscissa.
These delay-dependent fluorescence traces are characterized by

\begin{itemize}
  \item Low fluorescence yields at negative pump-probe delays.
  \item Increase of the fluorescence yield as the nozzle temperature is lowered to 12~K, which corresponds to a mean droplet radius $R\approx 9$~nm, and a decrease for lower nozzle temperatures corresponding to larger droplets.
  \item A step-like rise in the fluorescence yield when the XUV pump and NIR probe pulses arrive simultaneously, and a peaking fluorescence yield around a delay of 200~fs.
  \item A marginal drop in fluorescence yield for pump-probe delays between 0.2~ps and $\sim1$~ps; this drop is more pronounced for smaller droplets.
  \item Most notably, an exponential decrease of the fluorescence yield on a long timescale (1-10~ns).
\end{itemize}


As the pump-probe delay turns positive, the XUV pump pulse resonantly excites He$^*$ prior to the NIR pulse.
Although some fraction of the He$^*$ are likely to leave the droplets in the course of ultrafast droplet relaxation,
\cite{kornilovFemtosecondPhotoelectronImaging2011, mudrichUltrafastRelaxationPhotoexcited2020} a large part of the He$^*$ remain bound to the droplet surface.
These He$^*$ are tunnel ionized by the NIR pulse and thus act as seeds to drive the nanoplasma ignition in the same way as dopant atoms,
due to their much lower ionization energy compared to that of ground-state He atoms.
Therefore, we observe enhanced XUV fluorescence yields at positive delays.
However, if the delay is long and the NIR pulse arrives after the He$^*$ have decayed or desorbed from the droplets,
the nanoplasma ignition is again inefficient and the fluorescence yield drops.

Excited He droplets relaxing to the 1s2s\,$^1$S atomic state eventually form He$_2^*$ excimers that decay via spontaneous emission.
\cite{joppienElectronicExcitationsLiquid1993,karnbachElectronicExcitationDecay1995,vonhaeftenDiscreteVisibleLuminescence1997}
However, this process is expected to occur on time scales $> 2.5$~ns~\cite{vonhaeftenElectronicallyExcitedStates2005,vonhaeftenSizeIsotopeEffects2011} and $\sim1.6~\mu$s,~\cite{mckinseyTimeDependenceLiquidhelium2003}
which is much longer than the time scale of the present experiment.
He$^*$ can also desorb from the droplets, although this is expected to be a minor relaxation channel given that the droplets studied here are rather large.
\cite{ziemkiewiczFemtosecondTimeresolvedXUV2014,ziemkiewiczUltrafastElectronicDynamics2015}
Thus, the decrease in the fluorescence yield mainly reflects the timescale of rICD of pairs of He$^*$.

As the nozzle temperature $T_0$ is lowered, on the one hand, the flux of He droplets out of the nozzle increases;
on the other, the average size of He droplets also increases.
Both a higher droplet flux and the larger droplet size contribute to producing more excited and ionized He atoms, resulting in an overall rise in fluorescence yield.
When lowering the nozzle temperature further into the super-critical expansion regime at $T_0\leq11$~K, where the mean droplet radius increases to $R\geq17$~nm, the droplet number density drops and the droplet beam becomes more collimated. Under these conditions, the fluorescence yield decreases again.
The fast depletion of He$^*$ by rICD mediated by the merging of adjacent bubbles forming around each He$^*$, which was discussed earlier~\cite{laforgeUltrafastResonantInteratomic2021},
is visible as a small drop in fluorescence yield between the peak of the signal at 0.2-0.3~ps and about 1 ps delay.
This fast dynamics is more pronounced for small droplets with $R=5$~nm, where the initial distance between two He$^*$ is short,
mainly limited by the droplet size.

\begin{figure}[h]
    \centering
    \includegraphics[width=8.3cm]{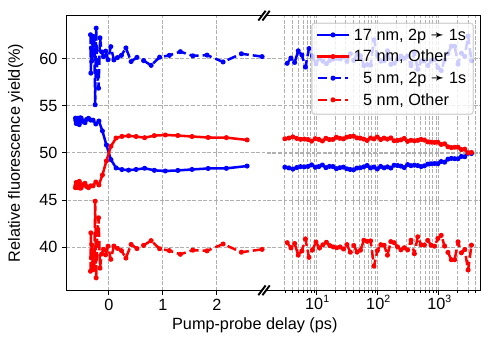}
    \caption{
        \label{fig:relative_yield}
        Relative fluorescence yield from the 2p state of excited He$^{+*}$ (blue) and all other states combined (red) as a function of pump-probe delay. Solid lines show the relative yield from large droplets with a mean radius $R\approx17$~nm, dotted lines are for small droplets with $R\approx5$~nm.
    }
\end{figure}

The fluorescence yield from different excited states of He$^{+*}$ also shows a droplet size-dependent behavior.
For a constant NIR pulse energy of 230~$\mu$J, the fluorescence from the 2p state decaying to the 1s state, compared to the fluorescence from higher excited states, is higher for small droplets and decreases for larger droplets.
Figure~\ref{fig:relative_yield} shows the relative fluorescence yields for different droplet sizes as a function of pump-probe delay.
The relative fluorescence from the 2p state is shown as blue symbols, while the sum of the relative yield of all higher states is shown in red.
The solid lines show the relative yields for large droplets ($R\sim~17$~nm), the dotted lines show the relative yields for small droplets ($R\sim~5$~nm).
It is well known that larger He droplets tend to be more efficiently avalanche ionized, thus forming a hotter and more highly charged nanoplasma~\cite{krishnanEvolutionDopantinducedHelium2012}.
The total amount of absorbed energy from the NIR probe, $W_{\text{abs}}$, is larger compared to small droplets. The same holds for optimum values of the pump-probe delay, in the present case 0.2-1000~ps.
Larger values of $W_{\text{abs}}$ lead to higher occupation of highly excited He$^{+*}$ states in the expanded nanoplasma. This trend is confirmed by classical molecular dynamics simulations, see Fig.~\ref{sfig:wabs_radius} in the supplementary material.
A more detailed analysis of the relative fluorescence yields at different experimental conditions is presented in the supplementary material.


To quantify the observed dynamics of the total fluorescence yield,
the fluorescence yield curves were analyzed by fitting them to a model that takes into account an exponential drop at short delays ($<1$~ps) and at long delays ($\gtrsim100$~ps). The model for the dynamics at short and long delays is given by
\begin{equation}
  \label{eq:fit_function_split}
    f_{s,l}(t) = \Theta(t) \exp\left(-\frac{t}{\tau_{s,l}}\right),
\end{equation}
where $\Theta(t)$ is the Heaviside step function and $\tau_{s,l}$ are the decay time constants of the exponential functions that model the decays at short and long delays.
Functions $f_{s,l}(t)$ were convolved with a Gaussian function that models the cross-correlation of pump and probe pulses.
The resulting convolved functions are given by
\begin{equation}
  \label{eq:fit_convolve}
  \mathcal{F}_{s,l}(t) = \exp\left[ \frac{1}{2} \left( \frac{\delta}{\tau_{s,l}} \right)^{2} - \frac{t}{\tau_{s,l}} \right] \erfc\left[ \frac{1}{\sqrt{2}} \left( \frac{\delta}{\tau_{s,l}} - \frac{t}{\delta}\right) \right].
\end{equation}
Here, $\delta$ is the standard deviation of the cross-correlation between pump-probe pulses.
The final fit function is taken as the weighted sum of each convolved exponential decay function,
\begin{equation}
  \label{eq:fit_function}
  \mathcal{F}(t) = A \mathcal{F}_s(t) + B \mathcal{F}_l(t).
\end{equation}
The fit yields the value $\delta \approx 95 \pm 15$~fs, corresponding to a cross-correlation FWHM between the pump and probe pulses of approximately 221~fs.
This is somewhat larger than the expected value of 112~fs estimated from the FWHMs of the XUV ($\sim$~50~fs) and NIR pulses ($\sim$~100~fs), possibly resulting from shot-to-shot fluctuations of the pulse durations and time jitter.
The resulting value for the fast rICD process, $\tau_s \approx 200 \pm 100$~fs, is in agreement with those reported by Laforge \textit{et al.}~\cite{laforgeUltrafastResonantInteratomic2021}.
Values for the long decay constant $\tau_l$ inferred from the fit are compiled in table \ref{tab:decay_time}.
These values depend on the droplet size and range from about 2~ns up to $\sim10$~ns for the largest droplets studied here.

\begin{table}[h]
  \small
  \centering
  \caption{
    \label{tab:decay_time}
    Comparison of the decay constants obtained from the experiment and from the Monte-Carlo molecular dynamics simulation.
  }
  \begin{tabular*}{0.7\textwidth}{cccc}
    \toprule
    Temperature (K) & Droplet radius (nm) & \multicolumn{2}{c}{Decay constant (ns)} \\
    \cmidrule{3-4}
     & & Experimental & Simulated \\
    \midrule
    14.5 & 5  & 1.9 & 1.96 \\
    14.0 & 7  & 3.7 & 3.30 \\
    12.0 & 9  & 5.5 & 5.35 \\
    {\color{gray} 11.0} & {\color{gray} 17} & {\color{gray} 8.6} & {\color{gray} 15.15} \\
    \bottomrule
  \end{tabular*}
\end{table}

The long time constant $\tau_l$ takes values of a few nanoseconds, and it increases with the droplet size.
From earlier time-dependent density functional theory (TDDFT) studies,~\cite{lehtovaaraEfficientNumericalMethod2004,barrancoHeliumNanodropletsOverview2006, ancilottoDensityFunctionalTheory2017} we know that within $< 250$~fs following the excitation of the droplet to the 1s2p\,$^1$P state, it relaxes to a state correlating to the 1s2s\,$^1$S state of the He atom. Subsequently, within $\sim 1$~ps, a void bubble forms around the excited He$^*$ atom on which the excitation localizes.
The bubbles forming around adjacent He$^*$ either merge into one, leading to fast rICD,~\cite{laforgeUltrafastResonantInteratomic2021}, or they are expelled to the droplet surface.~\cite{mudrichUltrafastRelaxationPhotoexcited2020}
When the bubbles burst at the droplet surface, the released He$^*$ are either ejected away from the droplet~\cite{kornilovFemtosecondPhotoelectronImaging2011}
or, for larger droplets, remain bound to the droplet surface.~\cite{benltaiefChargeExchangeDominates2019}

Due to the superfluid nature of He droplets, embedded atoms and molecules have previously been found to freely move through the droplets.~\cite{grebenevSuperfluiditySmallHelium41998, sheppersonLaserInducedRotationIodine2017, brauerCriticalLandauVelocity2013}
Therefore, we assume that surface-bound He$^*$ roam about the He droplet surface freely without experiencing any friction as long as their velocity stays below a critical value, analogously to the frictionless motion in the droplet interior.~\cite{brauerCriticalLandauVelocity2013}
This free motion is interrupted only when two He$^*$ atoms approach each other closely enough for their orbitals to overlap, at which point decay via rICD occurs.
Therefore, the rICD time constant is determined by the roaming motion of pairs of He$^*$ atoms at the droplet surface.
The depletion of the population of He$^*$ by rICD consequently leads to quenching of the fluorescence emission.

\subsection{Numerical simulations and discussion}

\begin{figure}[h]
  \centering
  \includegraphics[width=9cm]{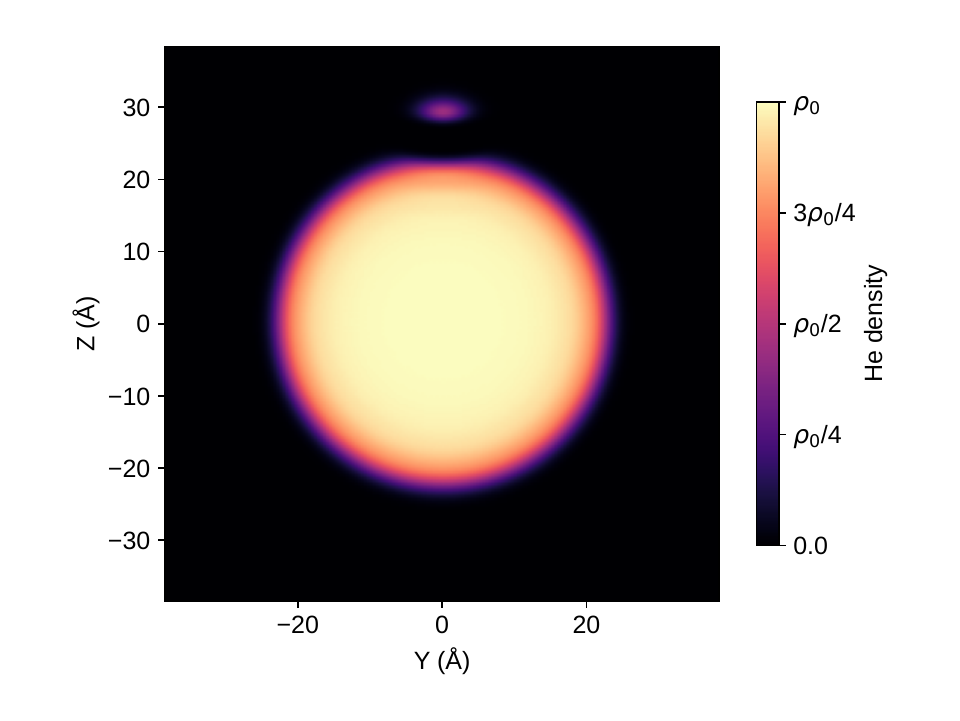}
  \caption{
    \label{fig:dft_simulation}
    Helium density functional theory (DFT) simulation showing the probability density of the He$^*$ wavefunction on a He nanodroplet of radius $R=2.22$~nm consisting of $10^3$ He atoms.
    The He$^*$ atom is bound to the droplet surface with a binding energy of $\sim9.7$~K, and its center of mass is located at 22.95~nm from the droplet center.
    The color bar shows the density of the He droplet in units of $\rho_0$, where $\rho_0$ is the density of bulk liquid He at zero temperature and pressure.
  }
\end{figure}

To assess our conjecture of He$^*$ roaming freely about the droplet surface and decaying by rICD upon their encounter, we performed a set of numerical simulations. First, we explored the equilibrium structure of a He$^*$ atom bound to a He nanodroplet.
To this end, a static $^4$He density functional theory ($^4$He-DFT) simulation was performed for a pure He droplet containing 1000 He atoms ($R=2.22$~nm) supporting one He$^*$ atom on its surface. The He$^*$ was treated as a quantum impurity due to the low mass. As explained in Ref.~\cite{ancilottoDensityFunctionalTheory2017}, where details about the method can be found, $^4$He-DFT has proven to be the best compromise between accuracy and the ability to treat a large number of atoms.
The He$^*$ in the 1s2s\,$^1$S state was described by a wave function and the rest of the droplet by its He density, both expanded on the same 3-dimensional Cartesian grid.
The equilibrium configuration was found by solving the coupled Euler-Lagrange equations arising from functional variation of the equation expressing the total energy as a functional of the He density.
The He$^*$-droplet interaction was taken as a sum of pairwise He$^*$-He potentials using the He$^*$(1s2s\,$^1$S)-He form of Fiedler and Eloranta.~\cite{fiedlerInteractionHeliumRydberg2014}

The simulations were performed using the free and open-source 4He-DFT-BCN-TLS code~\cite{pi4HeDFTBCNTLSComputer2024} with the Orsay-Trento functional.~\cite{dalfovoStructuralDynamicalProperties1995}
The resulting equilibrium configuration for He$^*$(1s2s\,$^1$S)@He$_{1000}$ is represented in Fig.~\ref{fig:dft_simulation} as a 2-dimensional cut in a polar plane.
He$^*$ is bound to the droplet surface $\sim2.95$~nm~away from the center of mass of the droplet with a binding energy of $\sim9.7$~K ($\sim0.84$~meV). In response, the droplet surface forms a shallow dimple underneath the He$^*$.
Thus, the simulation confirms that the He$^*$ atom can be bound to the surface of a He droplet.

\begin{figure}[h]
  \centering
  \includegraphics[width=9cm]{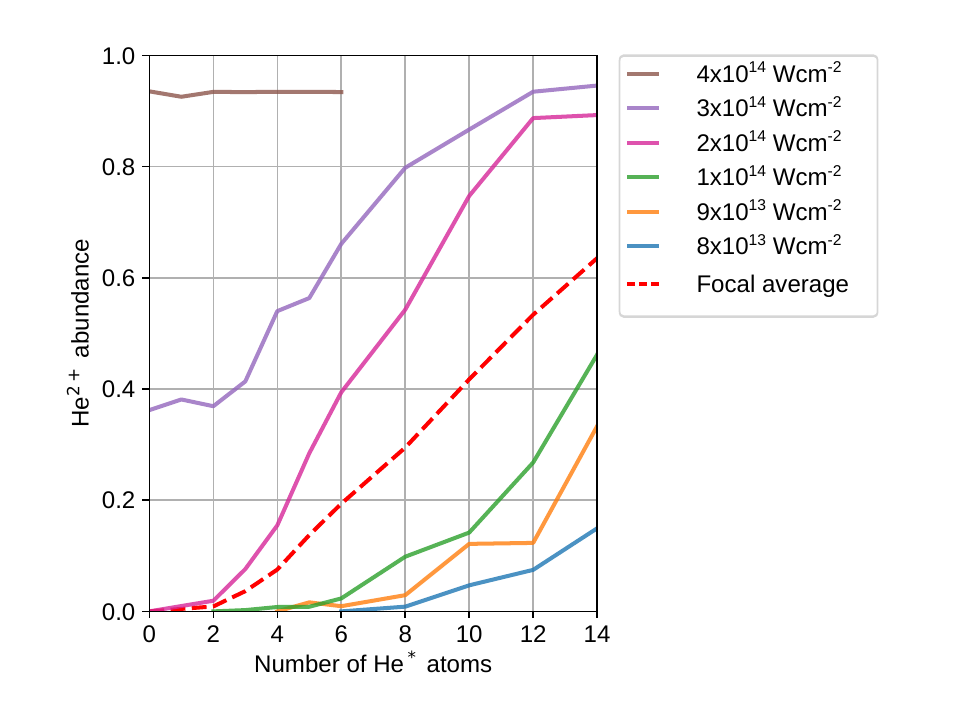}
  \caption{
    \label{fig:md_simulation}
    MD simulation of the He$^{2+}$ ion abundance as a function of the number of excited He$^*$ atoms attached to a He nanodroplet when the NIR pulse arrives.
    Different colors represent different intensities of the NIR pulse.
    The dashed red line represents the focal average over NIR intensities from $8\times 10^{13}$ to $2\times10^{14}$~Wcm$^{-2}$.
    From the focal averaged He$^{2+}$ abundance, it is clear that only a few He$^*$ atoms are needed as seeds to ignite a nanoplasma in the droplet.
  }
\end{figure}

Second, to assess the He$^*$ capacity to facilitate the ignition of a nanoplasma in a He nanodroplet,
we performed classical molecular dynamics (MD) simulations similar to those in earlier works.
\cite{medinaLonglastingXUVActivation2023,heidenreichChargingDynamicsDopants2017,heidenreichEfficiencyDopantinducedIgnition2016}
The He$^*$ were modeled by assigning them the excitation energy of 20.6 eV of the 1s2s\,$^1$S state, lowering their ionization energy by the corresponding amount.
These simulations were initialized by randomly placing a small number $n^*$ of He$^*$ inside a He nanodroplet containing 2171 He atoms ($R=2.6$~nm).
The evolution of this system under irradiation by an NIR pulse of 100~fs duration and various pulse peak intensities was computed until 220~fs after the NIR pulse maximum, when the ionization avalanche converged.
For each $n^*$ and pulse peak intensity, a set of 100 trajectories with different random positions of the He$^*$ was simulated and the abundance of He$^{2+}$ dications, which are the precursors of the experimentally observed nanoplasma fluorescence, was averaged over the trajectory set.
Figure~\ref{fig:md_simulation} shows the He$^{2+}$ trajectory-set abundances as a function of $n^*\leq14$ for various pulse peak intensities (solid curves).
Depending on the NIR pulse intensity, either $\gtrsim6$ He$^*$ ($I=10^{14}$~Wcm$^{-2}$) or  $\gtrsim2$ He$^*$ ($I>2\times10^{14}$~Wcm$^{-2}$) are needed to trigger ignition of a nanoplasma.
At intensities $I\ge3\times10^{14}$~Wcm$^{-2}$, even pure He nanodroplets that do not contain any He$^*$ are efficiently avalanche ionized.

For a better comparison with the experimental data, the dashed red curve shows the He$^{2+}$ ion abundance averaged over various intensities according to the intensity distribution in the volume determined by the spatial overlap of the XUV and the NIR beams.
The focal averaging involves the intensities that contribute to the He$^{2+}$ abundance, up to $I\leq2\times10^{14}$~Wcm$^{-2}$. For this intensity distribution, the probability of avalanche ionization of pure He nanodroplets containing no He$^*$ excitations is negligible as in the experiment for NIR pulse energies $<230~\mu$J [Fig.~\ref{fig:pump_probe}~(a)].
Clearly, at NIR pulse intensities close to the threshold for avalanche ionization of the pure He nanodroplet ($\sim3\times10^{14}$~Wcm$^{-2}$), exciting just a few He$^*$ already leads to nanoplasma ignition in some trajectories.
This confirms that resonant excitation is another efficient route to activation of He nanodroplets for subsequent NIR-induced nanoplasma ignition, similar to photoionization of the nanodroplets that led to the formation of stable He$^+_n$ `snowball complexes'.~\cite{medinaLonglastingXUVActivation2023}
However, in contrast to those earlier experiments, in which the probability of nanoplasma ignition remained constant over many nanoseconds, the activation by resonant excitation decays on the nanosecond timescale.
The focally averaged He$^+$ abundance (not shown in Fig.~\ref{fig:md_simulation}) is by a factor of 6 smaller than that of He$^{2+}$.
Details of the MD simulation method, including the focal averaging procedure and the simulation results, are presented in the Supplemental Material.

\begin{figure*}[ht!]
  \centering
  \includegraphics[width=0.8\textwidth]{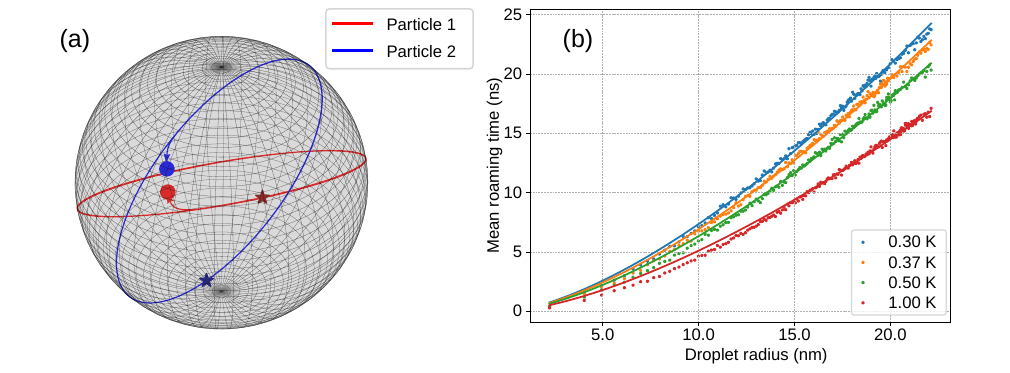}
  \caption{
    \label{fig:roam}
    (a) Illustration of two excited He$^*$ atoms roaming on the surface of a He droplet as simulated by a classical trajectory calculation.
    Atoms 1 and 2 are initialized at random positions marked by the symbols $\filledstar$
    and velocities randomly generated from a Maxwell-Boltzmann distribution at a temperature of 0.37~K; we neglect the weaker He-He$^*$ interaction and retain only the stronger He$^*$-He$^*$ interactions.
    The colored symbols $\medbullet$ indicate the endpoints of the trajectories. Both particles start their trajectories with constant velocities along the droplet's surface, following circular orbits on the sphere until they reach a short interatomic distance where they mutually attract each other and subsequently decay via rICD.
    (b) Mean roaming times of He$^*$ atoms before they meet and decay by rICD as a function of droplet radius; each value is the result of averaging over $10^5$ Monte-Carlo trajectories.
    Different colors show different initial velocity distributions given by the indicated temperatures.
    The simulated curves are fitted to a simple power-law, and in the case of $T=0.37$~K, the best fit is deduced so that the time constant scales as $\tau = 0.21R^{3/2}$.
  }
\end{figure*}

To rationalize the long decay dynamics observed in the experiment,
in a third step, we modeled the motion of He$^*$ atoms along the surface of a nanodroplet, assuming that two colliding He$^*$ decay by rICD.
To this end, classical trajectories of two He$^*$ were simulated using the Monte Carlo method;
the initial positions of the atoms were chosen randomly from a 3D uniform distribution of coordinates on the surface of the sphere.
The magnitude of the initial velocity was randomly generated according to a Maxwell-Boltzmann distribution for the He droplet temperature of 0.37~K.
The initial velocity vector of each He$^*$ was randomly oriented in a plane tangential to the surface at its initial position.
Previous experiments, in which the temperature of an atom or molecule attached to a He nanodroplet was measured, showed that the dopant atom or molecule fully thermalized to the droplet temperature $T=0.37$~K.~\cite{hartmannRotationallyResolvedSpectroscopy1995,aubockTripletStateExcitation2007}
Therefore, it appears probable that also the translational degree of freedom of a particle attached to a He nanodroplet thermalizes to the nanodroplet temperature.~\cite{toenniesSuperfluidHeliumDroplets2004}
Another plausible value would be the critical Landau velocity for frictionless motion of a molecule through superfluid He.~\cite{landauTheorySuperfluidityHelium1941, brauerCriticalLandauVelocity2013}
It turns out that for He$^*$, the thermal velocity, $v_\mathrm{th} = \sqrt{8 k_\mathrm{B} T / (\pi m_\mathrm{He})} = 44$~m/s has a similar value at $T = 0.37$~K as the critical Landau velocity, $v_c \approx 60$~m/s.
Using the velocity Verlet integrator,~\cite{verletComputerExperimentsClassical1967} Newton's equation of motion was solved numerically for each atom,
taking into account only the attractive force acting between the He$^*$ resulting from to the He$^*$-He$^*$ pair potential for 1s2s\,$^1S$ states,~\cite{fiedlerInteractionHeliumRydberg2014} and neglecting the interaction between the He$^*$ and the droplet.
The code \textit{Sphere roaming} is available under an open-source license.~\cite{sishodiaSphereRoamingProgram2025}

Figure~\ref{fig:roam}~(a) illustrates one sample trajectory of a pair of He$^*$ atoms orbiting around the sphere representing the He nanodroplet.
The randomly chosen initial positions of the atoms are marked with the symbols $\filledstar$ and the final positions are marked with $\medbullet$.
The attractive force acting between the He$^*$ has almost no effect as long as the distance between the He$^*$ exceeds $\sim 3$~nm.
Therefore, unless both particles were initialized at a distance $\leq 3$~nm, they follow circular orbits around the droplet,
deviating from these orbits only when coming close to each other.
When the atoms approach each other to less than 0.6~nm, the simulation is stopped, and the time it took the atoms to collide was recorded.
For each droplet size $10^5$ trajectories were simulated.

The resulting mean roaming time as a function of droplet radius $R$ is shown in Fig.~\ref{fig:roam}~(b).
Remarkably, at $T=0.37$~K, the mean roaming time follows a simple power law $\tau\propto R^{3/2}$.
This scaling lies between $\propto R$, which would be expected for direct collisions between He$^*$ roaming along the identical one-dimensional closed orbit, and $\tau\propto R^2$, which would be expected for diffusive motion of the He$^*$ exploring the entire surface area of the droplet.
Our simulation of He$^*$ moving along closed circular orbits, which, however, are randomly oriented with respect to each other
and each He$^*$ has a different velocity according to the thermal distribution, can be viewed as an intermediate case between the two scenarios described above.
Therefore, the observed $\tau\propto R^{3/2}$-scaling appears highly plausible.

Assuming only two He excitations occur in each droplet,
the roaming time was calculated as an average over the log-normal droplet size distribution for a given mean droplet size
sampled with $10^7$ values.
The resulting time constants $\tau_l$ match the experimental observations except for the case of large droplets with a mean radius $R=17$~nm, see table~\ref{tab:decay_time}.
A potential source of disparity between the experimental and simulated values lies in the determination of droplet sizes, particularly for larger droplets,
as the average droplet size was inferred from the literature rather than directly measured.
Moreover, larger droplets have a finite probability of accommodating more than one pair of He$^*$ per droplet, which reduces the roaming time.
Note that for large droplets, $\tau_l$ seems to approach the lifetime of the He$_2^*$ excimer, $<10$~ns.~\cite{mckinseyTimeDependenceLiquidhelium2003}
However, the lack of any signature of rICD involving He$_2^*$ in this and previous experiments~\cite{benltaiefSpectroscopicallyResolvedResonant2024} leads us to preclude He$_2^*$ from playing any role in the current experiment.

A similar but distinct process to rICD is exciton-exciton annihilation (EEA) occurring in different classes of nanoscale and extended condensed-phase systems, such as molecular aggregates and crystals, as well as certain photosynthetic pigments.
At high excitation densities, EEA proceeds by fusion of two excitons.
In this process, energy transfer takes place between the two excitons, leading to the annihilation of one exciton.~\cite{sunaKinematicsExcitonExcitonAnnihilation1970}
EEA is driven by purely electronic energy transfer and does not involve the mechanical motion of the molecules themselves, except for local vibrations that provides the energy match needed for resonance between donor and acceptor sites.
It can occur by transfer of electronic excitation energy from one excited molecule to another by direct F{\"o}rster transfer (via resonant dipole–dipole coupling) or Dexter transfer (via orbital overlap, corresponding to the rICD discussed here); in diffusive or multistep transfer, energy is transferred to a unexcited molecule followed by exciton annihilation.~\cite{fennelExcitonexcitonAnnihilationDisordered2015}
At high excitation densities, direct EEA initially dominates. As the excitation density decreases and the average distance between excitons increases, diffusive EEA predominates.
These exciton transfer mechanisms, while similar to the rICD process, involve two transitions between discrete energy levels and thus require resonance conditions for energy conservation. In contrast, in rICD the accepting molecule is ionized with the emitted electron carrying away the excess energy, thus offering more flexibility in conserving energy.
Low photon fluence and large droplet size reduce the He$^*$ concentration, thus favoring the roaming rICD mechanism that resembles diffusive (F{\"o}rster or Dexter) exciton transfer.
However, diffusive transfer occurs via exciton hopping within a static molecular assembly, whereas the analogous process in He droplets relies on the ballistic motion of excited He$^*$ atoms over large distances, enabled by the superfluid nature of He nanodroplets.

\section{Conclusion}
\label{sec:conclusion}

In summary, we studied the evolution of resonantly excited He nanodroplets using XUV fluorescence as a novel probe for the dynamics occurring on various time scales. Our experiment, backed by a set of simulations, demonstrates that resonant excitation of He nanodroplets by an XUV pulse efficiently activates the droplets for subsequent avalanche ionization and nanoplasma formation by a subsequent NIR pulse, even for a small number of excitations per droplet.
The most abundant XUV fluorescence emission from the nanoplasma was measured for transitions from the lowest excited levels to the ground state of the He$^+$ ion.
The fluorescence yields showed a slight drop at pump-probe delays between 0.2~ps and $\sim1$~ps, reflecting the fast decay of excited states by rICD mediated by the merging of bubbles formed around He$^*$ excitations.
More prominently, the fluorescence yield slowly drops on a time scale of several nanoseconds.
This slow dynamics was rationalized by the decay of pairs of droplet-bound excited He atoms by rICD, based on the detection of characteristic electrons and backed by numerical simulations.
Given the low concentration of He$^*$ excitations in each nanodroplet, the He$^*$ emerge to the nanodroplet surface, orbit around the nanodroplet, and eventually decay by rICD upon their encounter.
This process can be seen as an extreme case of rICD driven by nuclear motion, in contrast to most other previously reported instances of rICD where pure electronic couplings have been the main driving mechanism.~\cite{jahnkeInteratomicIntermolecularCoulombic2020}

Further experiments are planned that will resolve pump-probe delay-dependent rICD electron spectra, both for the given scenario of resonant excitation and for other rICD schemes such as He droplet excitation by photoelectron impact,~\cite{benltaiefEfficientIndirectInteratomic2023} photo-double excitation,~\cite{bastianObservationInteratomicCoulombic2024} and simultaneous photoionization and excitation.~\cite{shcherbininInteratomicCoulombicDecay2017}
Bright and tunable laser-based high-harmonic sources, which are now becoming readily available,~\cite{jurkovicovaBrightContinuouslyTunable2024} are particularly well suited for such experiments as  clusters and nanoparticles have large total absorption cross sections in the XUV range owing to their sheer size;
this allows us to study the ultrafast dynamics of collective excitations~\cite{laforgeUltrafastResonantInteratomic2021, jurkovicovaBrightContinuouslyTunable2024} and multiple photoionization processes~\cite{medinaLonglastingXUVActivation2023} with excellent time resolution, revealing complex interactions of nanoscale matter with ionizing radiation that may have implications for a range of related fields, including radiochemistry and radiation biology.

\ack{
    M.M. acknowledges financial support by the Novo Nordisk Foundation (grant no. NNF23OC0085401). L.B.L acknowledges support by the Villum foundation via the Villum Experiment grant No. 58859.
    SRK thanks Dept. of Science and Technology, Govt. of India, for support through the DST-DAAD scheme and Science and Eng. Research Board. A.H. gratefully acknowledges financial support from the Basque Government (project IT1584-22) as well as computational and manpower support of the DIPC computation center. S.R.K. and K.S. acknowledge the support of the Scheme for Promotion of Academic Research Collaboration, Min. of Edu., Govt. of India, and the Institute of Excellence programme at IIT-Madras via the Quantum Center for Diamond and Emergent Materials. S.R.K. gratefully acknowledges support of the Max Planck Society's Partner group programme. N.S. and S.R.K. gratefully acknowledge CEFIPRA (Indo-French Centre for the Promotion of Advanced Research) for generous support.
    The research leading to this result has been supported by the COST Action CA21101 ``Confined Molecular Systems: From a New Generation of Materials to the Stars (COSY)''.
    The authors thank the staff of the ELI Beamlines Facility, a European user facility operated by the Extreme Light Infrastructure ERIC,
    for their support and assistance.
}

\bibliographystyle{rsc}
\bibliography{lib}

\newpage

\appendix

\renewcommand{\thefigure}{S\arabic{figure}}
\setcounter{figure}{0}

\section{Molecular dynamics simulations}
\label{ssec:md_simulations}
The goal of the molecular dynamics (MD) simulations in this work is to explore how many electronically excited He atoms have to be generated in a droplet by the XUV pump pulse so that the NIR probe pulse triggers (``ignites'') an ionization avalanche, as well as to simulate the produced He$^{+}$, He$^{2+}$ ion yields.

The general features of the MD simulation method for the interaction of a cluster with the electric
and magnetic field of a linearly polarized NIR Gaussian laser pulse was described earlier.
\cite{heidenreichIonEnergeticsElectronrich2012, heidenreichEfficiencyDopantinducedIgnition2016, heidenreichExtremeIonizationXe2007, ammosovTunnelIonizationComplex1986}
In short, all nuclei and nanoplasma electrons are treated classically, starting with a cluster of neutral atoms. Electrons enter the MD simulation when the criteria for tunnel ionization (TI), classical barrier suppression ionization (BSI) or electron impact ionization (EII) are met. The criteria for TI, BSI and EII are checked at each atom at every MD time step,
using the local electric field at the atoms as the sum of the laser electric field and the contributions from all ions and electrons of the cluster.
Instantaneous TI probabilities are calculated by the Ammosov-Delone-Krainov formula,
\cite{ammosovTunnelIonizationComplex1986, ilkovIonizationAtomsTunnelling1992}
and EII cross sections by the Lotz formula,~\cite{lotzEmpiricalFormulaElectronimpact1967}
taking the ionization energy with respect to the atomic Coulomb barrier in the cluster.
\cite{fennelHighlyChargedIons2007}
Interactions between ions are described by Coulomb potentials, electron-ion
and electron-electron interactions by smoothed Coulomb potentials.
Interactions involving neutral atoms are disregarded except for a Pauli repulsive potential
between electrons and neutral He atoms.
Accordingly, the current simulation model cannot account for excited state nuclear dynamics
which would require the incorporation of at least He$^*$-He as well as He-He pairwise potentials.
Only from the instant of the first ionization on, the simulation model can describe nuclear
and electron dynamics.

The role of the electronic excitation is to activate the droplet for the subsequent ionization avalanche by the NIR probe pulse.
The activation consists of reducing the effective ionization energy from 24.6~eV for ground state He by
the electronic excitation energy to 4~eV for excited state He$^*$.
While in the experiment He is excited into the 1s2p\,$^1$P state and quickly relaxes to the 1s2s\,$^1$S state with $\lesssim$~0.5~ps,
\cite{mudrichUltrafastRelaxationPhotoexcited2020,laforgeRelaxationDynamicsExcited2022}
in our simulations we directly prepare the He$^*$ atoms in their 1s2s\,$^1$S state,
corresponding to the excitation energy of 20.6 eV.
Simulations are carried out for a fixed number $n^*$ of He$^*$ atoms, $n^*$ being a fixed given value for a trajectory.
Since excited state dynamics cannot take place in our simulation model and $n^*$ is a predetermined value,
the XUV pulse parameters are irrelevant as long as the electronic excitation is sufficiently separated from the NIR probe pulse in time.

We prepare the $n^*$ He$^*$ atoms randomly distributed over the droplet 300 fs prior to the NIR pulse maximum, which is sufficiently long given the intensity FWHM of the NIR probe pulse of 100~fs.
For each value of $n^*$, a set of 100-200 trajectories is simulated, each with different random sites of the He$^*$ atoms.
We restrict our simulations to the abundances of bare ion charges He$^+$ and He$^{2+}$, that is three-body recombination (TBR) which is the necessary prerequisite for the experimentally observed fluorescence via the formation of excited-state ions like the hydrogen-like He$^{+*}$ ion, is not considered.
Thus, by simulating bare ion charge abundances we obtain an indirect measure for the ability of the nanoplasma to fluoresce, assuming that the bare ion yield and the fluorescence yield are proportional to each other.
Classical TBR (involving two electrons and the target ion) occurs automatically in classical trajectory simulations but would require long trajectories for letting the nanoplasma expand and cool.
Since the goal of these simulations is to merely confirm the experimental observation that weak resonant excitation of He nanodroplets leads to their activation for subsequent nanoplasma ignition, we refrain from a more thorough modeling of the long-term dynamics that determines their fluorescence emission.
The trajectories are propagated until 220~fs after the NIR pulse maximum at which point the ionization avalanche has fully developed.

For a better comparison with the experiment,
the simulation results are averaged over the intensity profile in the NIR focal spot.
A corresponding focal averaging over the XUV intensity profile is not necessary,
since the trajectories are run for fixed given values of $n^*$.
For the focal averaging of the NIR focal spot we start from a 3D intensity profile of a Gaussian beam~\cite{meschedeOpticsLightLasers2007}

\begin{equation}
I(r,z) = I_{\text{max}}\frac{w_0^2}{w(z)^2}\exp\left( - \frac{2r^{2}}{w(z)^{2}} \right),
\end{equation}
where $I_{\text{max}}$ is the intensity in the center of the focal spot, $w(z)=w_0\sqrt{1+z^2/z_0^2}$ is the $z$-dependent beam radius, $w_0$ is the beam radius in the beam waist, $z$ is the propagation direction of the photons and $r$ the radial coordinate perpendicular to $z$.
For the Rayleigh length $z_0=\pi w_{0}^{2}/\lambda$ the ratio is $z_{0}^{\text{XUV}} / z_{0}^{\text{NIR}}$~=~54, taking the focal spot size $w_0^2$ of the XUV ($\lambda$~=~57~nm) to be 64$\times$47~$\mu$m$^2$ and focal spot size of the NIR ($\lambda$~=~796~nm) to be 18$\times$43~$\mu$m$^2$.
Thus, the XUV intensity profile is considerably more elongated than that of the NIR beam. Since in the experiments both beams are aligned nearly collinearly, the XUV intensity profile cuts through the 3D NIR volume along the $z$-axis and activates only droplets along a narrow channel.
Accordingly, focal averaging reduces approximately to a 1D averaging over a Lorentzian intensity profile along the $z$-axis,
\begin{equation}
I(z) = I_{\text{max}}\frac{z_{0}^{2}}{z_{0}^{2} + z^{2}}.
\end{equation}

The 1D focal averaging is carried out analogously to the procedures for 2D and 3D intensity profiles described in Ref.~\cite{heidenreichKineticEnergyDistribution2011}.
To this end, MD simulations are carried out for selected NIR pulse peak intensities between $8\times10^{13}$ and $2\times10^{14}$~Wcm$^{-2}$,
i.e., from the lowest intensity at which droplets contribute to the ion yield up to the highest sampled intensity at which no ignition occurs in absence of He$^*$.

The simulations are carried out mainly for the He$_{2171}$ droplet, whereas the smallest average droplet size considered in the experiment is $10^4$ He atoms.
Some simulations are also performed for droplet sizes up to $10^4$ atoms to discuss the droplet size dependence of the ion signals.
The structures of the droplets is assumed to be a FCC lattices with a He-He distance of 3.6~\AA.~\cite{peterkaPhotoionizationDynamicsPure2007}

\begin{figure}
    \centering
    \includegraphics[width=0.6\textwidth]{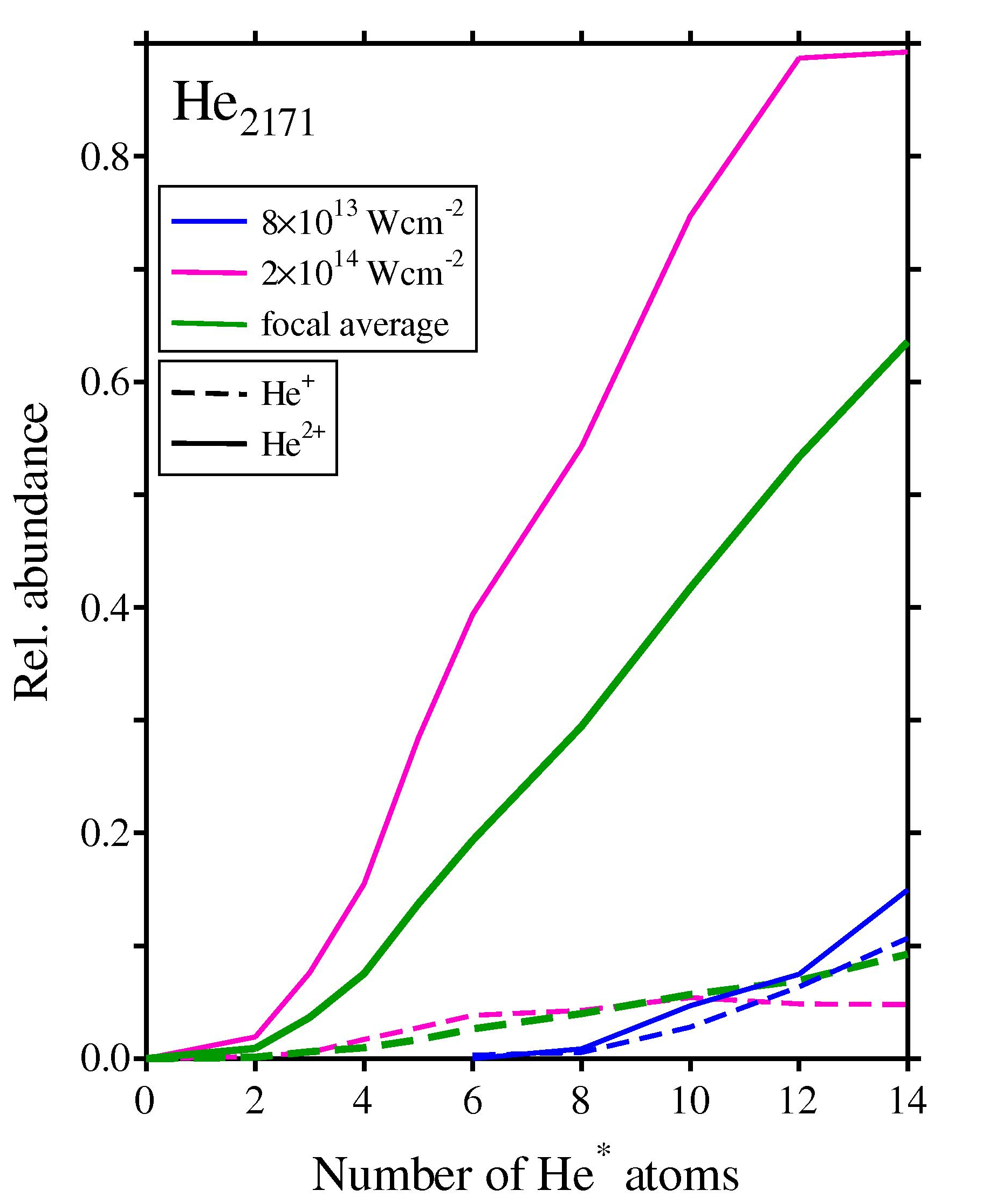}
    \caption{
        He$^+$ (dashed curves) and He$^{2+}$ (solid curves) abundances for the He$_{2171}$ droplet as a function of the number of He$^*$ atoms.
        The green curves represent the focally averaged results involving all intensities $8\times 10^{13}$~$\leq I \leq 2\times 10^{14}$~Wcm$^{-2}$
        presented in figure~\ref{fig:md_simulation} of the main text.
        Given are also the results for the single intensities $I = 8\times 10^{13}$ (blue) and $I =2 \times 10^{14}$~Wcm$^{-2}$ (magenta).
        \label{sfig:he2_abundance}
    }
\end{figure}

As stated in the main text, the focally averaged simulated He$^+$ abundance (normalized to the total number of He atoms in the droplet)
is by a factor of six lower than that of He$^{2+}$.
Figure~\ref{sfig:he2_abundance} exhibits the He$^+$ and He$^{2+}$ abundances.
Presented are the focally averaged results (green curves) as well as the results for the lowest
and highest single NIR pulse peak intensity involved in the focal averaging.
At the lowest intensity, $I = 8\times 10^{13}$~Wcm$^{-2}$ (blue curves),
the He$^+$ and He$^{2+}$ abundances are comparable,
while at the highest intensity, $I = 2\times 10^{14}$~Wcm$^{-2}$ (magenta curves),
He$^{2+}$ predominates by far.
As a general trend, for a fixed intensity the abundances of both He$^+$ and He$^{2+}$ increase with increasing $n^*$, because ignition takes place in more trajectories.
Further, the presence of more He$^*$'s tends to ignite the droplet at earlier times during the NIR pulse,
leading to a more efficient avalanche and thus the generation of more He$^{2+}$ per single trajectory.

\begin{figure}
    \centering
    \includegraphics[width=0.6\textwidth]{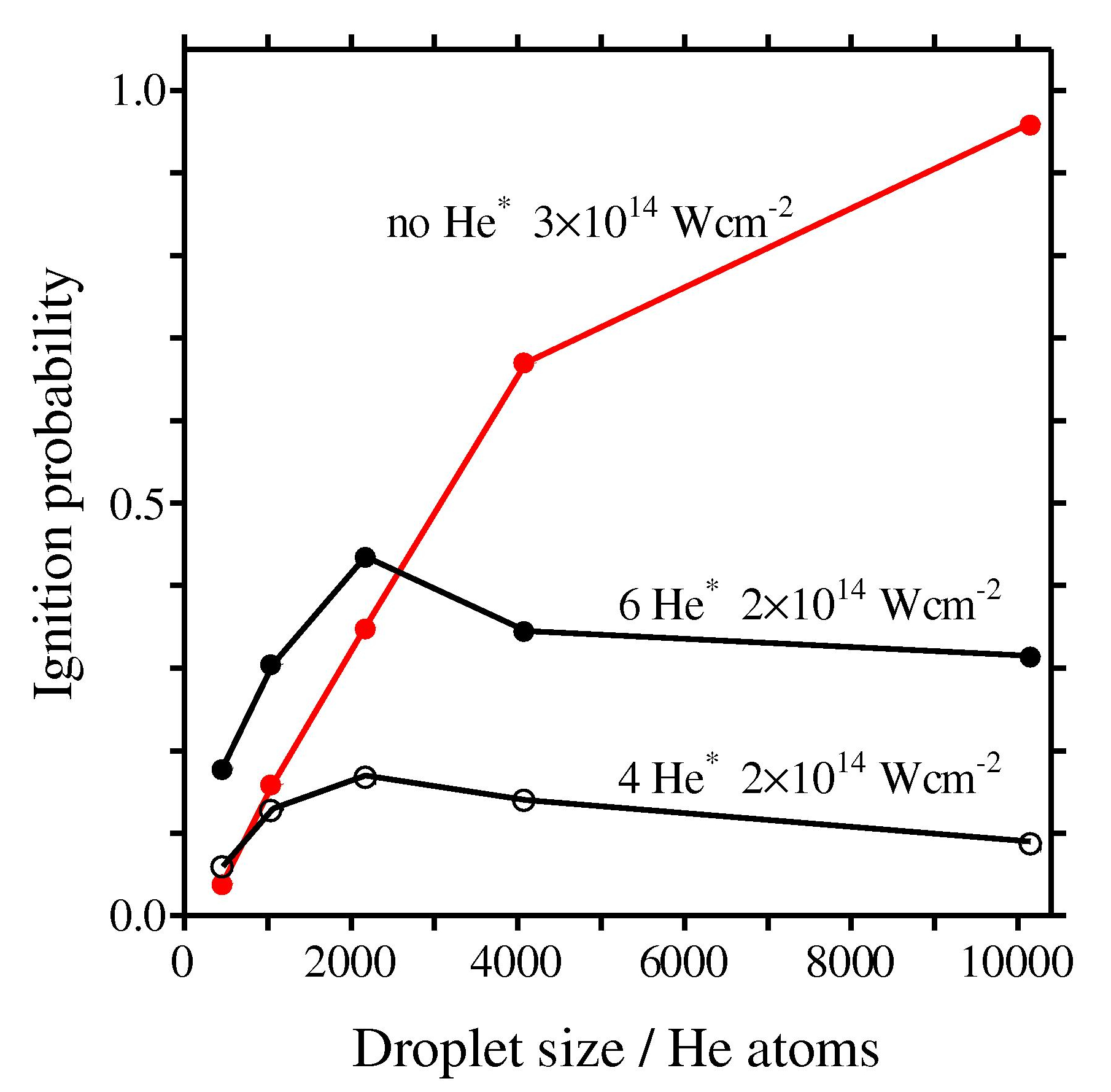}
    \caption{
        The droplet size dependence of the ignition probability, shown for the droplet sizes of 459, 1016, 2171, 4096, and 10149 He atoms.
        Given are three examples: $I = 3\times 10^{14}$ Wcm$^{-2}$, $n^*=0$ (red curve); and $I = 2\times 10^{14}$~Wcm$^{-2}$, $n^* =4$ and 6 (black curves).
        \label{sfig:dropsize_md}
    }
\end{figure}

Figure~\ref{sfig:dropsize_md} shows the droplet size dependence of the ignition probability for five droplet sizes up to $\approx 10^4$ atoms.
The ignition probability is defined as the fraction of trajectories of a set in which an ionization avalanche is observed.
As an ignition criterion we take a minimum threshold of 200 ionizations in the droplet.
The threshold value is uncritical as the ionization avalanche was found to propagate through all the droplet once the threshold is passed.
At $I = 3\times 10^{14}$~Wcm$^{-2}$ and without the presence of He$^*$'s (red curve),
a strong droplet size dependence is exhibited, as the cumulative tunnel ionization probability,
which grows with the number of atoms in the droplet, reaches a level at which a single tunnel ionization somewhere inside the droplet occurs,
is sufficient at this intensity to trigger an avalanche.
While for the He$_{2171}$ droplet, for which the simulation results of Figure~\ref{fig:md_simulation} in the main text are carried out, the ignition probability amounts to only 0.35, for the average droplet size range $\geq 10^4$ atoms ignition occurs with near certainty.
For $I = 2\times 10^{14}$~Wcm$^{-2}$,
ignition probabilities are presented for $n^* = 4$ and 6.
After a steep rise with a weak maximum at $2\times 10^3$ atoms,
the probabilities show a weak droplet size dependence,
only slightly decreasing towards larger droplet sizes.
The cause for the weak maximum at $2\times 10^3$ atoms is unknown.
One may speculate about two opposite trends which may result in a maximum for the ignition probability:
(1) With increasing droplet size, the electrons released from the He$^*$'s have a longer path through the droplet and consequently have more opportunities to cause secondary ionizations.
(2) With increasing droplet size the probability decreases that two He$^*$'s are generated in a close neighborhood in the droplet.
The He$^+$ ions generated from the He$^*$'s
and by secondary ionizations have then fewer opportunities to collectively lower the Coulomb barriers at neutral He atoms in a specific region in the droplet.

The weak droplet size dependence of the ignition probability at $I = 2\times 10^{14}$~Wcm$^{-2}$ suggests that the simulation results for the He$_{2171}$ droplet are applicable to the much larger droplets in the experiment.
Further and not shown in figure~\ref{sfig:dropsize_md},
the absence of tunnel ionization for $n^* = 0$ even for droplets of $10^4$ atoms extends the validity of $2\times 10^{14}$~Wcm$^{-2}$ as the maximum intensity for the focal averaging at least up to this droplet size.

\section{State-resolved fluorescence yields}
\label{ssec:relative_fluorescence_yield}

\begin{figure}[t]
    \centering
    \includegraphics[width=0.95\textwidth]{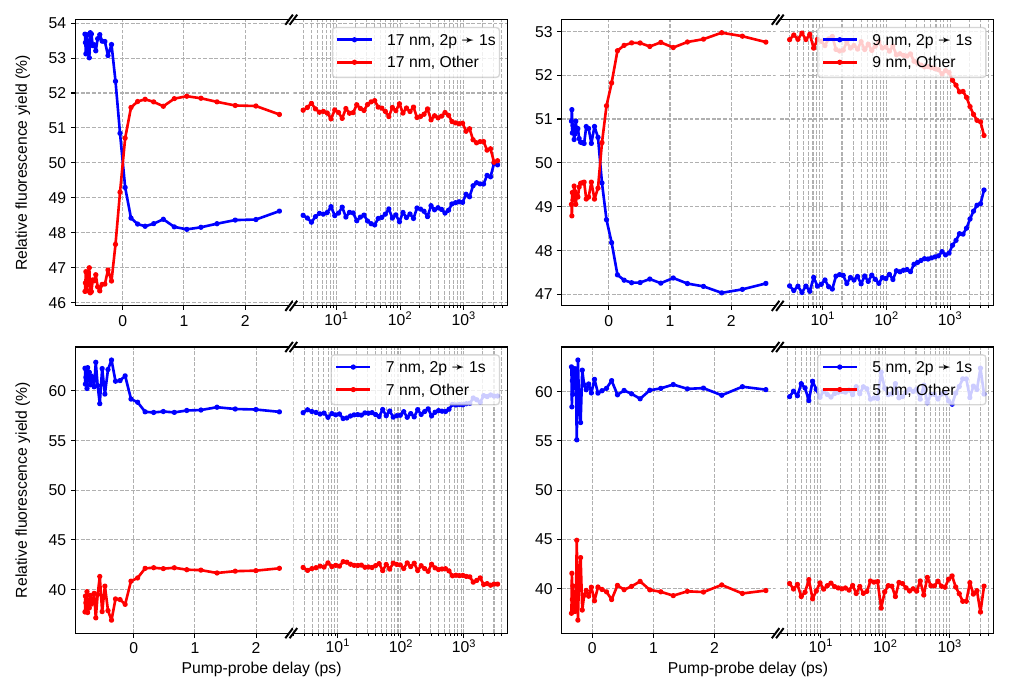}
    \caption{
        \label{sfig:relative_yield}
        Relative fluorescence yield as a function of pump-probe delay.
        The blue line indicates the relative fluorescence yield from the 2p$\to$1s state relative to the total fluorescence yield,
        while the red line shows the yield from all higher states.
        At negative delays, fluorescence arises from nanoplasma created solely by the NIR pulse.
        At positive delays, nanoplasma is generated by the tunnel ionization of excited He$^*$ atoms by the NIR pulse.
        For larger droplets, the fluorescence from higher states is stronger than that from the lowest fluorescing 2p state.
    }
\end{figure}

Figure~\ref{sfig:relative_yield} shows the relative fluorescence yield for different droplet sizes as a function of the pump-probe delay.
Relative fluorescence yield from the 2p state of excited He$^{+*}$ is shown in blue color,
while the relative yield of all the other states combined is shown in red color.
At negative delays, fluorescence arises from nanoplasma created by inefficient ignition of the droplet solely by the NIR pulse,
which shows a higher fluorescence yield from the 2p$\to$1s state.
For large droplets, the fluorescence from higher states dominates over the fluorescence from the 2p$\to$1s state.

\begin{figure}
    \centering
    \includegraphics[width=0.8\textwidth]{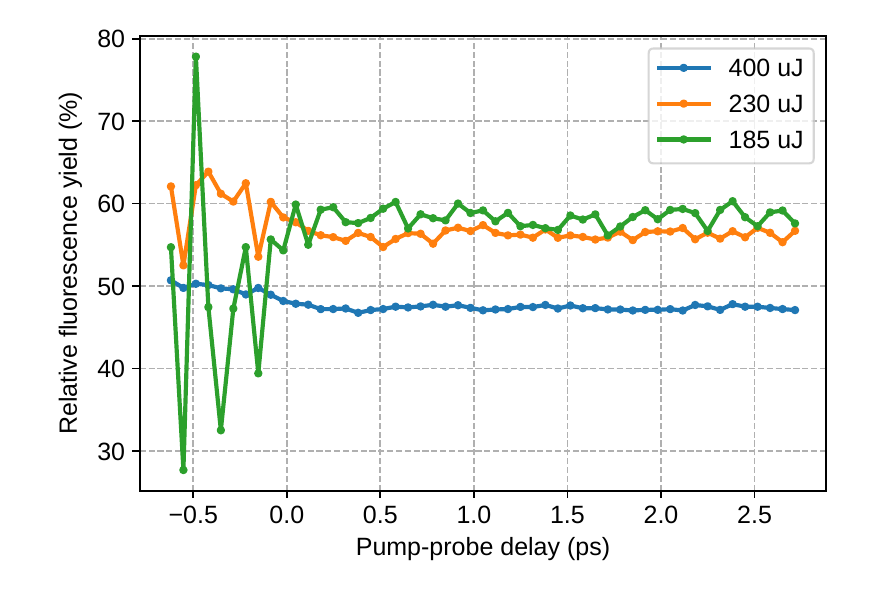}
    \caption{
        \label{sfig:relative_yield_intensity}
        Relative fluorescence yield of the 2p state to the total fluorescence yield as a function of pump-probe delay for different NIR pulse energies for the He droplet with a radius of 7~nm,
        which contains 30,000 He atoms.
        For high NIR pulse energies, the relative fluorescence yield from the 2p state drops,
        thereby increasing the relative fluorescence yield from higher states.
    }
\end{figure}

Figure~\ref{sfig:relative_yield_intensity} shows the fluorescence yield from the lowest fluorescing 2p state relative to the total fluorescence for different NIR pulse energies. For higher pulse energies, the He droplet absorbs more energy, accumulating more quasifree electrons and ions.
This, in turn, leads to a higher ion-electron recombination rate, leading to the occupation higher-lying states of He$^{+*}$, thus increasing the relative fluorescence yield from higher states and decreasing that of the 2p$\to$1s transition.
See figure~\ref{fig:pump_probe}~(a) for the total fluorescence yield as a function of pump-probe delay.

\begin{figure}
    \centering
    \includegraphics[width=0.8\textwidth]{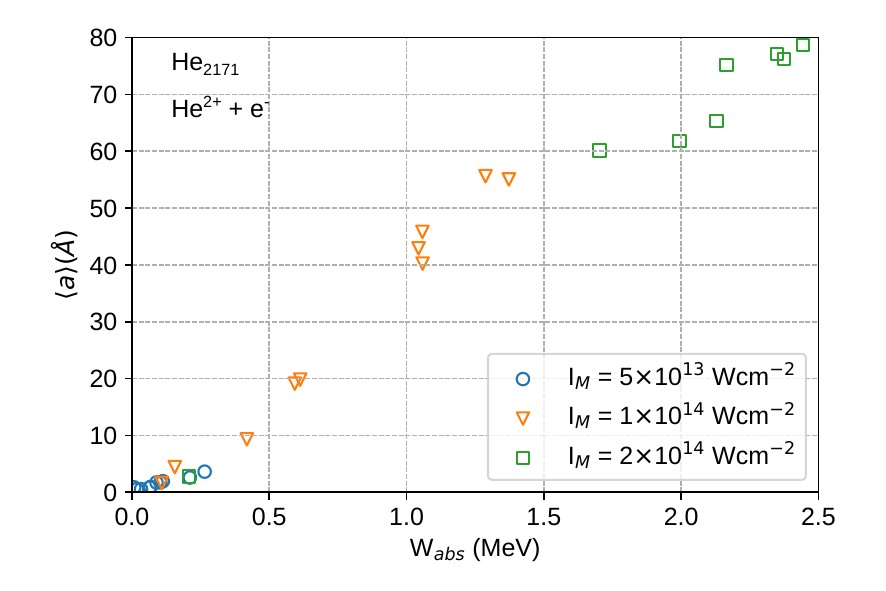}
    \caption{
        \label{sfig:wabs_radius}
        The average semimajor axis ($\langle a \rangle$) of the electrons in their Kepler ellipses around their central He$^{2+}$ dications
        as a function of laser energy absorbed by the nanoplasma, $W_{abs}$.
        Every data point in the figure corresponds to one trajectory, with $\langle a \rangle$ being the semimajor axis of the classical electron orbits averaged over all He$^{2+}$-electron systems of the trajectory.
    }
\end{figure}

Figure~\ref{sfig:wabs_radius} shows a characterization of the classical orbits of He$^{2+}$-electron system formed by three-body-recombination in the course of MD trajectories of He$_{2171}$ droplets.
Shown are the average semimajor axis values $\langle a \rangle$ of the electrons in their elliptical Kepler orbits around the central He$^{2+}$ ions.
Every data point in the figure corresponds to one trajectory, with $\langle a \rangle$ being the average over all He$^{2+}$-electron systems of the trajectory.
$\langle a \rangle$ is plotted against the laser energy $W_{\text{abs}}$ absorbed by the nanoplasma.
Given are the results for three MD trajectory sets for different NIR pulse peak intensities and numbers of He’s,
each set consisting of about 10 trajectories, which are extended to 3~ps to allow the nanoplasma to expand:
$5\times10^{13}$ Wcm$^{-2}$ with $n^*= 20$ (blue), $1\times10^{14}$ Wcm$^{-2}$ (orange) and $2\times10^{14}$ Wcm$^{-2}$ (green), the latter two data sets for $n^* = 14$.
The data points depend nearly linearly on $W_{\text{abs}}$.
For a given constant pulse peak intensity and a fixed number of He$^*$’s, $W_{\text{abs}}$ varies considerably,
being determined by the random instant at which the ionization avalanche is triggered during the laser pulse, as shown by Heidenreich \textit{et al.}~\cite{heidenreichChargingDynamicsDopants2017}
We note that the simulated electron orbits are quite excentric with excentricities ranging between 0.2 and 0.9 (not shown).

When this ignition instant occurs early, the nanoplasma absorbs more laser energy.
The trend of increasing $\langle a \rangle$ with increasing $W_{\text{abs}}$ is in qualitative agreement with the experimental observation that the fluorescence yield from higher-lying states of He$^{+*}$ increases with increasing laser intensity, see figure~\ref{sfig:relative_yield_intensity}. Naturally, the classical simulation does not allow for assignment of specific quantum numbers. Nevertheless, to relate to the experiment, $\langle a \rangle$ can be compared to the expectation value of the radii of He$^{+*}$ orbitals, $\langle r\rangle=3a_0n^2/4$ (assuming the orbital angular momentum quantum number $\ell=0$), that is $\langle r\rangle=1.6$~\AA~for $n=2$, $\langle r\rangle=3.6$~\AA~for $n=3$, $\langle r\rangle=6.4$~\AA~for $n=4$, etc. Here, $a_0=0.53$~\AA~is the Bohr radius.

The clear correlation of $\langle a \rangle$ with $W_{abs}$ can be rationalized as follows:
When a nanoplasma ignites, nearly all He atoms are doubly ionized in the ``inner-ionization'' state at short times;
however, when the nanoplasma expands, it rapidly cools and electrons recombine with the ions. If the plasma is highly energetic (large $W_{abs}$ values, high electron energies, He atom fully doubly ionized), then the nanoplasma expands fast and electrons recombining with He$^{2+}$ dications are trapped in high Rydberg states.
At lower $W_{\text{abs}}$, the expansion is slower, allowing for collisions between electrons and He ions in the expanding cloud.
This causes electrons to be de-excited into lower-lying states, as the effective cross sections for electron-He$^{+*}$ collisions are large for high principal quantum numbers of He$^{+*}$ states, $n$.~\cite{ralchenkoElectronimpactExcitationIonization2008}
In case of incomplete ionization of He atoms in the nanoplasma at small $W_{abs}$ values, the average electron kinetic energy is lower.
Thus, electron-He collisions by trend lead to impact excitation of lower states of He$^{*}$ and He$^{+*}$. In contrast, in energetic plasmas electron-He collisions rapidly lead to complete double ionization of He atoms and bound excited states are populated exclusively by electron-ion recombination in the expansion phase of the nanoplasma.
In our experiments, probably both mechanisms of populating excited states in He$^{*}$ and He$^{+*}$ contribute.

This interpretation is supported by our experimental findings for variable droplet size and variable NIR intensity, see figures~\ref{sfig:relative_yield} and \ref{sfig:relative_yield_intensity}. Both for small droplets (\textit{i.\,e.} droplet radius $R=5$, 7~nm) and for low NIR intensity (\textit{i.\,e. } at pulse energy 185~$\mu$J), $W_{\text{abs}}$ is low and, consequently, the relative fluorescence yield from the lowest excited state $n=2$ is highest.

\end{document}